\documentclass[aps,reprint,superscriptaddress,nofootinbib,longbibliography,notitlepage,floatfix]{revtex4-2}
\usepackage{graphicx}
\usepackage{tabularx}
\usepackage{amsmath}
\usepackage{amstext}
\usepackage{amssymb}
\usepackage{xfrac}
\usepackage[colorlinks,citecolor=blue]{hyperref}

\usepackage{graphicx}
\usepackage{amsmath}
\usepackage{amstext}
\usepackage{amssymb}
\usepackage{amsfonts}
\usepackage{longtable,booktabs}
\usepackage{hyperref}
\usepackage{url}
\usepackage{subfigure}
\usepackage{dsfont}
\usepackage{booktabs}
\usepackage{amsbsy}
\usepackage{dcolumn}
\usepackage{amsthm}
\usepackage{bm}
\usepackage{esint}
\usepackage{multirow}
\usepackage{hyperref}
\usepackage{cleveref}
\usepackage{amsmath}
\usepackage{amsfonts}
\usepackage{amssymb}
\usepackage{mathrsfs}
\usepackage{amsfonts}
\usepackage{amsbsy}
\usepackage{dcolumn}
\usepackage{bm}
\usepackage{multirow}
\usepackage{color}
\usepackage{extarrows}
\usepackage{datetime}
\usepackage{comment}
\usepackage[super]{nth}
\usepackage{tikz-cd}

\usepackage{makecell}
\usepackage{booktabs}
\usepackage{threeparttablex}

\usepackage{scalerel}
\usepackage{tikz}
\usetikzlibrary{svg.path}

\usepackage{xr}
\definecolor{orcidlogocol}{HTML}{A6CE39}
\tikzset{
	orcidlogo/.pic={
		\fill[orcidlogocol] svg{M256,128c0,70.7-57.3,128-128,128C57.3,256,0,198.7,0,128C0,57.3,57.3,0,128,0C198.7,0,256,57.3,256,128z};
		\fill[white] svg{M86.3,186.2H70.9V79.1h15.4v48.4V186.2z}
		svg{M108.9,79.1h41.6c39.6,0,57,28.3,57,53.6c0,27.5-21.5,53.6-56.8,53.6h-41.8V79.1z M124.3,172.4h24.5c34.9,0,42.9-26.5,42.9-39.7c0-21.5-13.7-39.7-43.7-39.7h-23.7V172.4z}
		svg{M88.7,56.8c0,5.5-4.5,10.1-10.1,10.1c-5.6,0-10.1-4.6-10.1-10.1c0-5.6,4.5-10.1,10.1-10.1C84.2,46.7,88.7,51.3,88.7,56.8z};
	}
}

\newcommand\orcidicon[1]{\href{https://orcid.org/#1}{\mbox{\scalerel*{
				\begin{tikzpicture}[yscale=-1,transform shape]
					\pic{orcidlogo};
				\end{tikzpicture}
			}{|}}}}

\hypersetup{
	colorlinks=magenta,
	linkcolor=blue,
	filecolor=magenta,
	urlcolor=magenta,
}
\def\Z{\mathbb{Z}}

\newtheorem*{theorem*}{Theorem}

\def\C{\mathcal{C}}

\def\P{\mathbf{P}}

\def\l{\mathbf{l}}
\def\p{\mathbf{p}}
\def\L{\mathbf{L}}

\def\D{\mathsf{D}}
\def\S{\mathsf{S}}
\def\M{\mathsf{M}}
\def\P{\mathsf{P}}

\def\Z{\mathbb{Z}}

\usepackage{extarrows}

\usepackage{datetime}

\usepackage{comment}

\usepackage{lineno}

\usepackage{booktabs}

\begin{document}
	
	
	\title{Fragmented Topological Excitations in Generalized Hypergraph Product Codes}
	
	
	
	\author{Meng-Yuan Li\orcidicon{0000-0001-8418-6372}}
	\affiliation{Institute for Advanced Study, Tsinghua University, Beijing, 100084, China}
	\email{my-li@mail.tsinghua.edu.cn}
	
	\author{Yue Wu}
	\affiliation{Institute for Advanced Study, Tsinghua University, Beijing, 100084, China}

 	\date{\today}

\begin{abstract}
	
	Product code construction is a powerful tool for constructing quantum stabilizer codes, which serve as a promising paradigm for realizing fault-tolerant quantum computation. Furthermore, the natural mapping between stabilizer codes and the ground states of exactly solvable spin models also motivates the exploration of many-body orders in the stabilizer codes. In this work, we investigate the fracton topological orders in a family of codes obtained by a recently proposed general construction. More specifically, this code family can be regarded as a class of generalized hypergraph product (HGP) codes. We term the corresponding exactly solvable spin models \textit{orthoplex models}, based on the geometry of the stabilizers. In the 3D orthoplex model, we identify a series of intriguing properties within this model family, including non-monotonic ground state degeneracy (GSD) as a function of system size and non-Abelian lattice defects. Most remarkably, in 4D we discover \textit{fragmented topological excitations}: while such excitations manifest as discrete, isolated points in real space, their projections onto lower-dimensional subsystems form connected objects such as loops, revealing the intrinsic topological nature of these excitations. Therefore, fragmented excitations constitute an intriguing intermediate class between point-like and spatially extended topological excitations. In addition, these rich features establish the generalized HGP codes as a versatile and analytically tractable platform for studying the physics of fracton orders.
	
\end{abstract}


\maketitle

\tableofcontents

\section{Introduction}
\label{sec:intro}
The connection between quantum stabilizer codes and topological matter has been a fruitful area of research, stimulating numerous advances in both fault-tolerant quantum computation and the understanding of topological orders~\cite{Gottesman1997,Dennis2002,Kitaev2003,Castelnovo2008,Bravyi2011,Chamon2005,Haah2011,Vijay2015,Vijay2016,Ma2018a,He2018,Weinstein2019,Kong2020,Rakovszky2024}. More specifically, a stabilizer code can be viewed as the ground state subspace of an exactly solvable spin model, whose Hamiltonian $H = -\sum_i S_i$ is a sum of mutually commuting operators $\{S_i\}$, termed stabilizers. A paradigmatic example is the 2D toric code model, where the topological ground state degeneracy underpins the fault tolerance of the encoded quantum information~\cite{Dennis2002,Kitaev2003}. As a systematic generalization of the 2D toric code, hypergraph product (HGP) codes thus offer valuable insights into the study of both quantum codes and topological orders~\cite{Tillich2009,Zeng2019a}. The HGP framework constructs quantum CSS codes taking a set of classical codes as input. The 2D toric code itself corresponds to the HGP of two 1D repetition codes, which can be viewed as a classical 1D Ising model from the perspective of many-body systems.

Mathematically, the HGP can be naturally interpreted in the language of chain complexes~\cite{Tillich2009,Zeng2019a,Bombin2007a}. Classical and quantum CSS codes correspond to chain complexes of length $2$ and $3$, respectively. The HGP procedure involves taking the tensor product of the chain complexes corresponding to the input codes, followed by selecting a segment of length $3$ from the resulting tensor product complex as the output code. Recently, a general code construction has been proposed by systematically decomposing and reconstructing the tensor product complexes, that significantly extends the scope of product codes~\cite{Wu2025}. A subclass of the general construction can be regarded as a generalized HGP protocol, where the direct summands of chain groups of the tensor product complexes are recombined to obtain a series of quantum CSS codes beyond the standard HGP codes (see Sec.~\ref{sec:hd_models}). A natural question then arises: do these generalized HGP codes host nontrivial many-body orders?

In this paper, we answer this question affirmatively. By investigating the physics of a series of generalized HGP codes with 1D repetition codes as inputs, we show that these codes realize a family of models exhibiting fracton orders. This family complements our previous work on fracton orders in general dimensions~\cite{Li2020,Li2021,Li2023}. Here, fracton order refers to a novel class of long-range entangled phases characterized by topological excitations with restricted mobility~\cite{Chamon2005,Haah2011,Vijay2015,Vijay2016,Nandkishore2019,Pretko2020,Zhou2022a}. This fascinating property positions fracton models as an exotic form of matter exhibiting nontrivial interplay between topology and geometry, which challenges our conventional understanding of quantum phases and holds potential for more robust quantum memories~\cite{Haah2011,Song2022}. Since the stabilizers in the model family considered in this paper are defined on orthoplexes of various dimensions, we term this class of codes \textit{orthoplex models}.

First, we investigate the 3D orthoplex model that realizes a Type-I fracton order. This model is a rotated version of the model that has been discussed in Ref.~\cite{Shirley2019,Fuji2019,Liu2023}, thus it inherits the two types of fundamental lineon excitations, denoted $\l^e$ and $\l^m$, respectively. A dipole composed of lineons of the same type is mobile within a 2D plane (constituting a planon) and is correspondingly denoted by $\p^e$ or $\p^m$. Surprisingly, we find that introducing a dislocation into this model realizes a non-Abelian lattice defect, which exchanges $\p^e$ and $\p^m$ when they are braided around it~\cite{Bombin2010, You2012a, You2019}. Furthermore, we noticed that as proved in Ref.~\cite{Wu2025}, this model exhibits a non-monotonic ground state degeneracy (GSD) as a function of system size, unlike the unrotated model~\cite{Shirley2019,Fuji2019,Liu2023}. Such exotic scaling of GSD shows the nontrivial dependence of fracton orders on geometry~\cite{Chamon2005,Haah2011,Vijay2015,Slagle2017a,Slagle2018}.

\begin{figure}[t]
	\centering
	\includegraphics[width=\columnwidth]{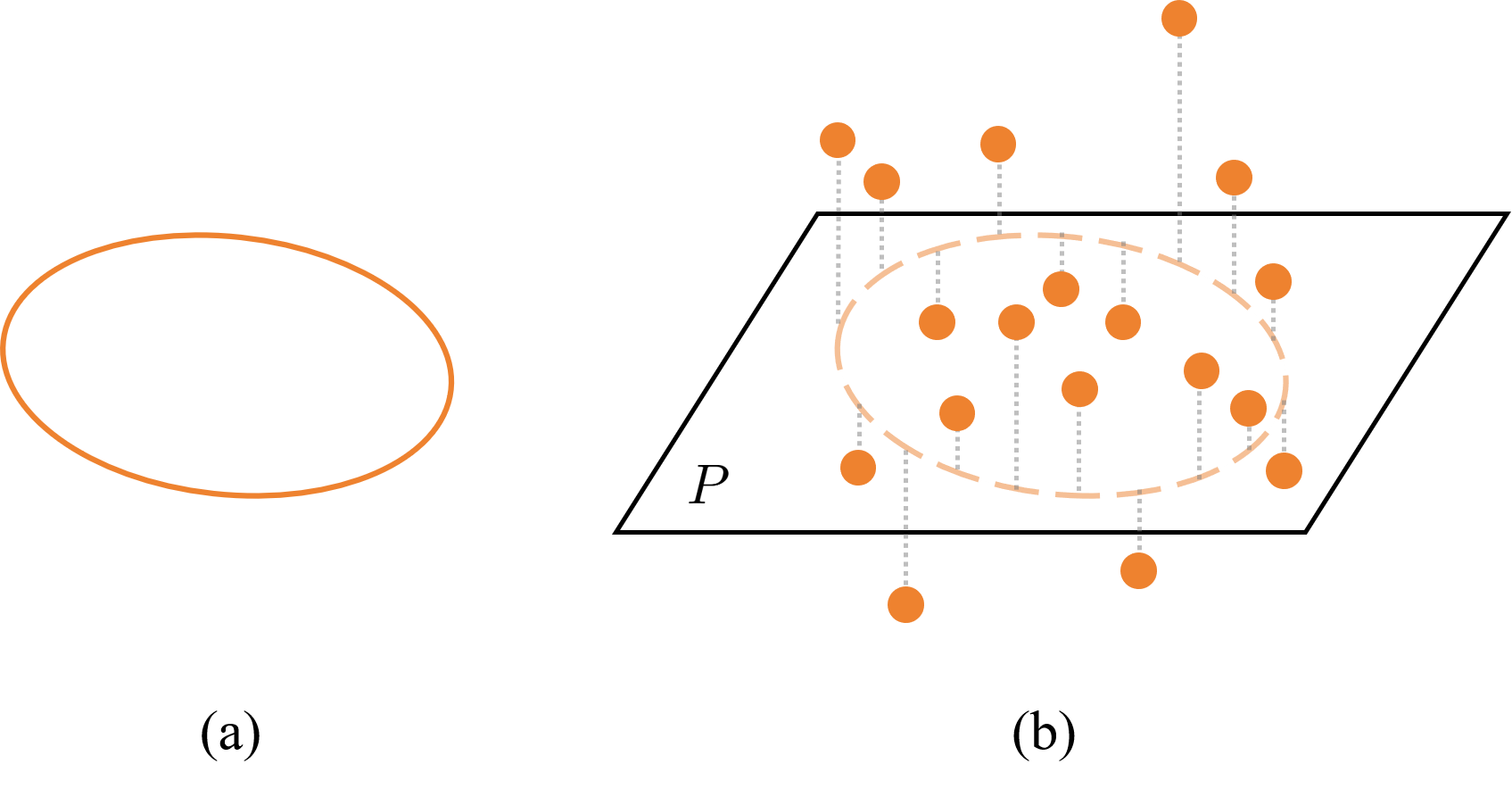}
	\caption{A schematic picture of fragmented topological excitations. In (a), for comparison, we show a standard loop excitations with orange color, where excitation energy is distributed along the loop. In (b), we show a fragmented loop excitation composed of point-like excitations scattered in a 3D space, while actually its projection onto the 2D plane $P$ forms a connected loop (the dashed orange circle), revealing its topological nature.}
	\label{fig:dis_loop}
\end{figure}

Moreover, we find \textit{fragmented topological excitations} in orthoplex models, that interpolate between point-like and spatially extended excitations. Conventional topological systems feature excitations that are either point-like or form contiguous, extended structures, such as loops and membranes~\cite{Hamma2005,Walker2012,Lan2018,Lan2019,Chan2018,Wen2018a,Wang2014,Jian2014,Jiang2014,Ning2016,Ning2018a,Ye2018,Wang2015a,Chen2016,Wang2016b,Putrov2017}. Such extended excitations are essential for the characterization of topological phases and are also an important ingredient for realizing self-correcting quantum codes~\cite{Dennis2002, Bombin2009, Bombin2015, Quintavalle2021, Higgott2023}. In the orthoplex models, however, we find that an intrinsically extended topological excitation can manifest as a collection of disconnected points; we thus term them ``fragmented''. As a concrete example, we investigate the 4D orthoplex model that hosts fragmented loop excitations. Such an excitation appears as a set of discrete point-like excitations, yet its projection onto a 2D plane forms a connected loop, revealing its underlying topological nature (see Fig.~\ref{fig:dis_loop}). Therefore, fragmented excitations serve as an intriguing intermediate type between point-like and spatially extended topological excitations.

This paper is organized as follows. In Sec.~\ref{sec:pre}, we briefly review fracton orders, and introduce useful notations for discussing (hyper)cubic lattices of arbitrary dimensions. In Sec.~\ref{sec:3d_model}, as a preliminary study, we analyze the 3D orthoplex model, demonstrating the existence of non-monotonic GSD and non-Abelian defects. In Sec.~\ref{sec:4d_model}, we introduce the concept of fragmented topological excitations, using the fragmented loop excitations in the 4D orthoplex model as a concrete example. Finally, in Sec.~\ref{sec:hd_models}, we review HGP codes and present the general construction of orthoplex models of arbitrary dimensions via a generalized HGP protocol. In Sec.~\ref{sec:summary}, we briefly discuss the implications of our results and potential future directions.

\section{Preliminaries}
\label{sec:pre}

\subsection{Review of fracton orders}
\label{subsec:rev_fracton}

This section briefly reviews the key features of fracton orders. Fracton orders represent a class of topological orders with long-range entanglement, distinguished by topological excitations with restricted mobility~\cite{Chamon2005,Haah2011,Vijay2015,Vijay2016,Nandkishore2019,Pretko2020,Zhou2022a}. For topological excitations, restricted mobility means that excitations at different spatial locations can be of distinction topological types~\cite{Wen2003,Kitaev2003,Vijay2015,Pai2019}. In other words, moving such an excitation along certain directions cannot be accomplished by a local operator; instead, it requires a non-local operation that acts on a macroscopic region of the system. From another perspective, such a movement realized by a local operator typically creates additional gapped excitations, rendering it energetically costly and contrary to the common expectation about movement of excitations. This stands in contrast to conventional topological orders, where point-like excitations such as anyons are fully mobile.  In Sec.~\ref{sec:3d_model}, we will demonstrate such mobility restriction using 3D orthoplex model as a concrete example.

Based on excitation mobility, fracton orders are broadly categorized into two types~\cite{Vijay2016}. Type-I fracton orders contain subdimensional particles, which are mobile within a subspace of the system, alongside completely immobile excitations termed fractons. In three spatial dimensions, prominent examples of subdimensional particles include lineons, which are mobile along one-dimensional lines, and planons, which are mobile within two-dimensional planes. The X-cube and checkerboard models exemplify Type-I fracton orders~\cite{Vijay2015,Vijay2016}. In contrast, Type-II fracton orders contain only completely immobile excitations (fractons). Haah's cubic code is a canonical example of Type-II fracton orders~\cite{Haah2011}.

The phenomenology of fracton orders extends beyond point-like excitations to spatially extended objects~\cite{Li2020,Li2021,Li2023}. The restrictions can apply not only to the mobility but also to the deformability of these extended excitations. For instance, a loop-like excitation may be restricted to move and deform only within a specific plane. Moreover, these objects can fuse to form even more intricate spatially extended excitations with non-manifold-like shapes, termed complex excitations in Ref.~\cite{Li2020}.

\subsection{Notations}

In this paper, as we mainly focus on (hyper)cubic lattice in arbitrary dimensions, it is beneficial to introduce a coordinate representation of lattices and cells wherein. In a $p$-dimensional hypercubic lattice, we assume the lattice constant to be $1$, and set the origin $(0,0,\cdots,0)$ at a vertex. Then, we can uniquely label a $q$-dimensional cell (a $q$-cell) by a $p$-tuple of coordinates. This tuple consists of $q$ half-integers, corresponding to the axes along which the cell extends, and $p-q$ integers, specifying its location. Here, a $q$-cell is a $q$-dimensional analog of a cube, for example, $0$-cell, $1$-cell and $2$-cell are respectively vertex, link and plaquette.

\begin{figure}[t]
	\centering
	\includegraphics[width=\columnwidth]{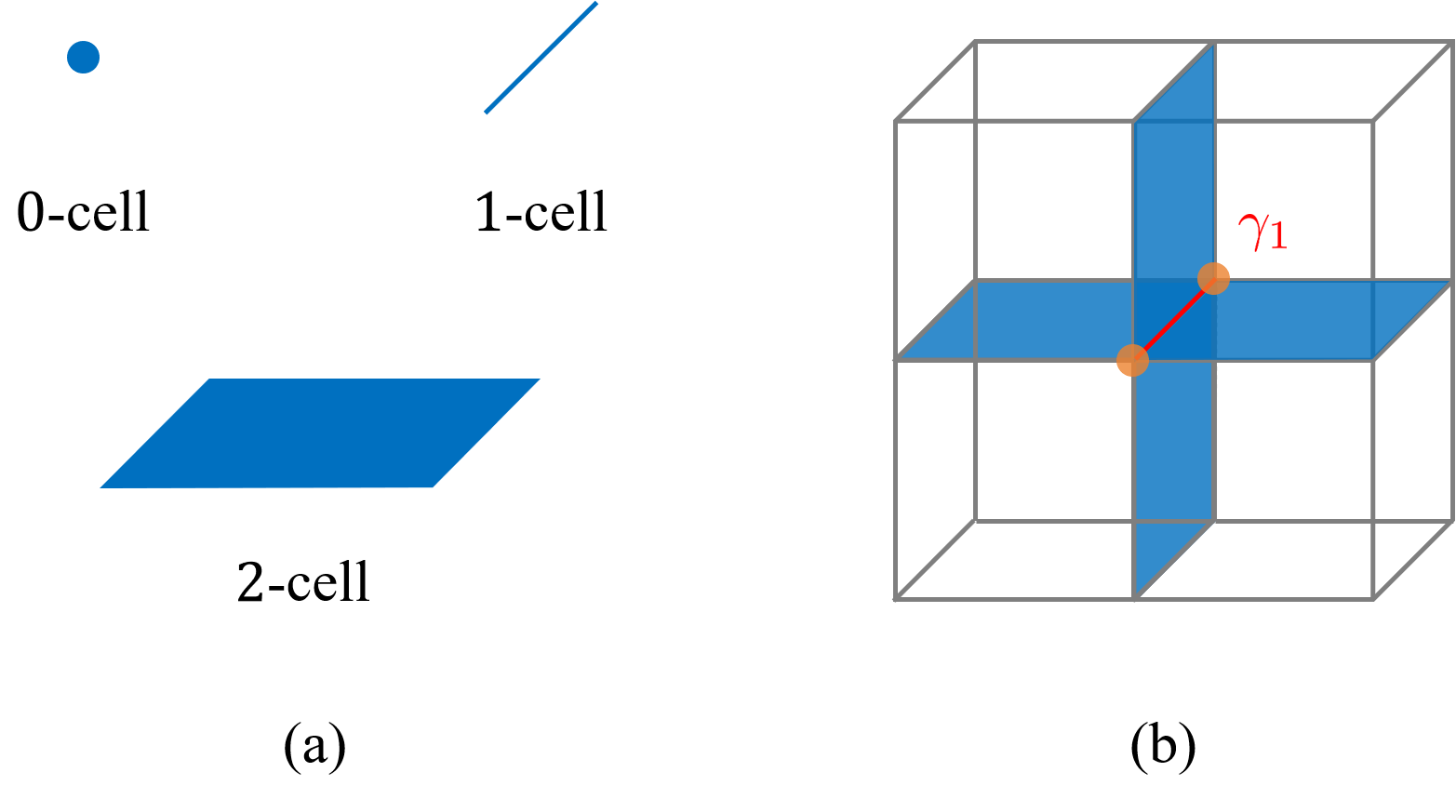}
	\caption{Cells in a 3D cubic lattice. In (a), we show a $0$-cell, a $1$-cell and a $2$-cell, all colored blue. In (b), we show a $1$-cell $\gamma_1$ (colored red) and the set of nearest cells of it $\Delta \gamma_1$, that is composed of two vertices (colored orange) and four plaquettes (colored blue).}
	\label{fig:cells}
\end{figure} 

In general, we use $\gamma_d$ to refer to a $d$-cell. When there is no ambiguity, we use $\gamma_d$ and its coordinate representation interchangeably. As a concrete example, in a 6D hypercubic lattice, we can write $\gamma_4 = (\frac{1}{2}, \frac{1}{2}, \frac{1}{2}, \frac{1}{2}, 0, 0)$ to define a $4$-cell centered at $(\frac{1}{2}, \frac{1}{2}, \frac{1}{2}, \frac{1}{2}, 0, 0)$. For simplicity, we also introduce the notation $\Delta \gamma_d$ to refer to the set of nearest neighbor cells of $\gamma_d$, which can be explicitly presented as:
\begin{align}
	\Delta \gamma_d \equiv \{\gamma_d \pm \frac{1}{2}\hat{x}_\mu | \mu=1, \dots, p\},
\end{align}
where $\hat{x}_\mu$ for $\mu=1, \dots, p$ are the basis vectors of the lattice. For example, given a link $\gamma_1 = (\frac{1}{2},0,0)$ in a 3D cubic lattice, we have $\Delta \gamma_1 = \{(0,0,0),(1,0,0),(\frac{1}{2},\frac{1}{2},0),(\frac{1}{2},-\frac{1}{2},0),(\frac{1}{2},0,\frac{1}{2}),(\frac{1}{2},0,-\frac{1}{2})\}$, that is a set of two vertices and four plaquettes (see Fig.~\ref{fig:cells}).

\section{3D Orthoplex Model}
\label{sec:3d_model}

In this section, we introduce the 3D orthoplex model, which provides a beneficial intuitive picture for the 4D model with fragmented topological excitations and general orthoplex models. This model, defined on a cubic lattice of size $L_x \times L_y \times L_z$, is a rotated version of the model investigated in Ref.~\cite{Shirley2019,Fuji2019,Liu2023}. We denote the three spatial directions as $x, y$ and $z$. Our construction via the generalized HGP imposes a rotation that results in a non-monotonic dependence of the ground state degeneracy (GSD) on the system sizes, that reflects the interplay between topology and geometry in fracton orders.

\begin{figure}[t]
	\centering
	\includegraphics[width=\columnwidth]{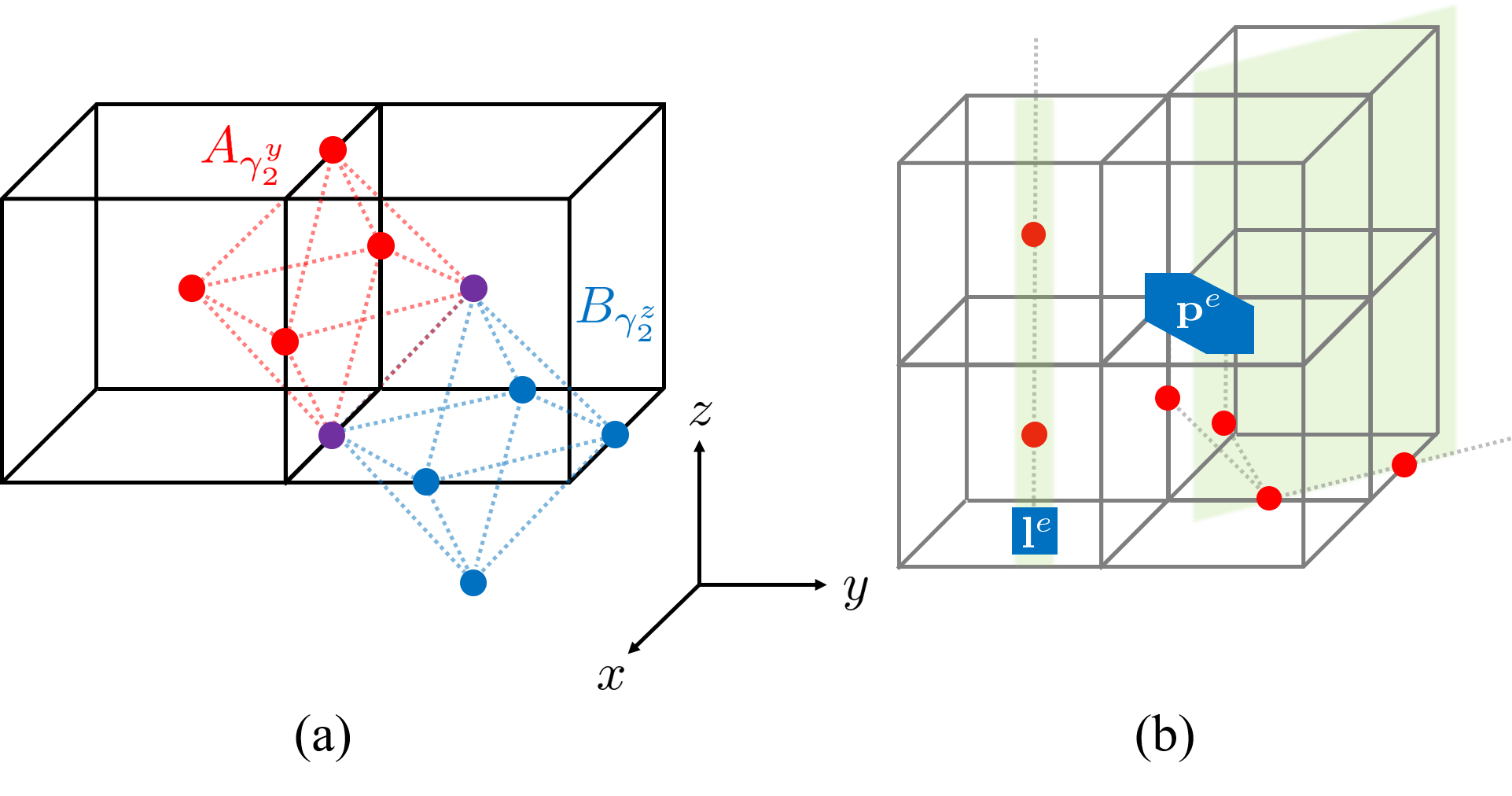}
	\caption{Hamiltonian and excitations in 3D orthoplex model. In (a), we show two representative Hamiltonian terms (i.e., stabilizers). Qubits involved in the $A$ term ($X$-stabilizer) and $B$ term are respectively colored red and blue, and the two qubits shared by the two terms are colored purple. In (b), we demonstrate a lineon $\l^e$ and a planon $\p^m$, each is generated by a string composed of Pauli $X$ operators denoted by the gray dotted line, with red filled circles specifying the qubits acted by the Pauli $X$ operators. For clarity, the mobile regions of the two excitations are schematically highlighted by the transparent green regions.}
	\label{fig:3dm+e}
\end{figure}

The model is specified by the following assignment of qubits and stabilizers (see Fig.~\ref{fig:3dm+e} (a)):
\begin{itemize}
	\item \textbf{Qubits}: A qubit is assigned to each link ($\gamma_1$) and each cube ($\gamma_3$).
	\item \textbf{$X$-stabilizers}: An $X$-stabilizer is assigned to each plaquette ($\gamma_2$) normal to the $x$- or $y$-axis.
	\item \textbf{$Z$-stabilizers}: A $Z$-stabilizer is assigned to each vertex ($\gamma_0$) and each plaquette ($\gamma_2$) normal to the $z$-axis.
\end{itemize}

The Hamiltonian of this model is a sum of commuting stabilizer operators:
\begin{align}
	H_\text{3D} = -\sum_{\gamma^s \in \Gamma_A} A_{\gamma^s} -\sum_{\gamma^s \in \Gamma_B} B_{\gamma^s},
\end{align}
where $\Gamma_A$ denotes the set of $\gamma_2^{x}$ and  $\gamma_2^{y}$, $\Gamma_B$ denotes the set of $\gamma_2^{z}$ and  $\gamma_0$, $\gamma_2^{\mu}$ denotes a plaquette normal to the $\mu$-axis ($\mu=x,y,z$). The stabilizer operators are defined as $A_{\gamma^s} = \prod_{\gamma^q \in \Delta \gamma^s} X_{\gamma^q}$ and $B_{\gamma^s} = \prod_{\gamma^q \in \Delta \gamma^s} Z_{\gamma^q}$, where the product runs over all qubit-bearing cells $\gamma^q$ nearest to the stabilizer cell $\gamma^s$. This stabilizer-qubit relationship gives rise to the model's characteristic orthoplex geometry: each stabilizer operator involves six qubits whose positions form the vertices of an octahedron (a 3D orthoplex) centered on the stabilizer. The $A$-type and $B$-type stabilizers are arranged in a layered structure: octahedra centered on planes with a half-integer $z$ coordinate correspond to $A$ stabilizers, while those on planes with an integer $z$ coordinate correspond to $B$ stabilizers.

As derived in Ref.~\cite{Wu2025}, the GSD of the 3D orthoplex model with periodic boundary conditions is
\begin{align}
	\label{eq:3d_gsd}
	\log_2 \text{GSD} = 4 \gcd(L_x, L_y).
\end{align}

This non-monotonic behavior with respect to system size is a hallmark of fracton order, also observed in models such as Chamon's code and Haah's cubic code~\cite{Chamon2005, Haah2011}. The GSD is determined by counting the independent logical operators. For this model, there are three types of logical Pauli operators (described here for the $X$-type, with the $Z$-type being analogous due to self-duality of stabilizers): 1) vertical string operators $W(\S^z) = \prod_{\gamma_1 \in \S^z} X_{\gamma_1}$ along a non-contractible straight loop in the $z$-direction; 2) diagonal string operators $W(\S^{x\pm y}) = \prod_{\gamma_1 \in \S^{x\pm y}} X_{\gamma_1}$ along a non-contractible diagonal loop along the direction of $x\pm y =0$; and 3) special operators $W(\P) = \prod_{\gamma_1 \in \P} X_{\gamma_1}$ supported on a planar region $\P$ whose shape is complicated and depends on the parity of $L_x$ and $L_y$ (see Ref.~\cite{Wu2025} for more details). This GSD stands in sharp contrast to the results for the model in Refs.~\cite{Shirley2019,Fuji2019,Liu2023} that differs from 3D orthoplex model by a rotation, where $\log_2 \text{GSD}$ grows linearly with $L_x$ and $L_y$.

\subsection{Topological excitations in 3D orthoplex model}

We now turn to the fundamental excitations. We first focus on excitations created by Pauli-$X$ operators, which violate the $B$ terms. A single violated $B$ term constitutes a lineon excitation, denoted $\l^e$, which is mobile only along the $z$-direction. Any attempt to move it within the $xy$-plane via a local Pauli-$X$ operator necessarily creates additional excitations, rendering such a process energetically costly.

While a single $\l^e$ is restricted, a dipole of lineons can acquire subdimensional mobility. Such a dipole, denoted $\p^e$, behaves as a planon. Consider a dipole $\p^e$ composed of two $\l^e$ excitations located at cells $\gamma_0^r$ and $\gamma_0^r + (\frac{1}{2},\frac{1}{2},0)$ within the $z=0$ plane (here superscript $r$ is for ``reference''). Applying a Pauli-$X$ operator to the qubit on the link $\gamma_0^r+ (\frac{1}{2},0,0)$ annihilates the original dipole and creates a new one, effectively translating $\p^e$ along the direction of diagonal line $x+y = 0$. Since each constituent lineon is mobile along the $z$-axis, the dipole $\p^e$ as a whole acts as a planon, mobile within the 2D plane spanned by the directions of $z$-axis and diagonal line $x+y = 0$. A pictorial demonstration of $\l^e$ and $\p^e$ excitations is given in Fig.~\ref{fig:3dm+e} (b). Due to the model's self-duality, excitations $\l^m$ and $\p^m$, created by Pauli-$Z$ operators violating $A$ terms, exhibit completely analogous properties.

\subsection{Non-Abelian defects in 3D orthoplex model}

\begin{figure}[t]
	\centering
	\includegraphics[width=\columnwidth]{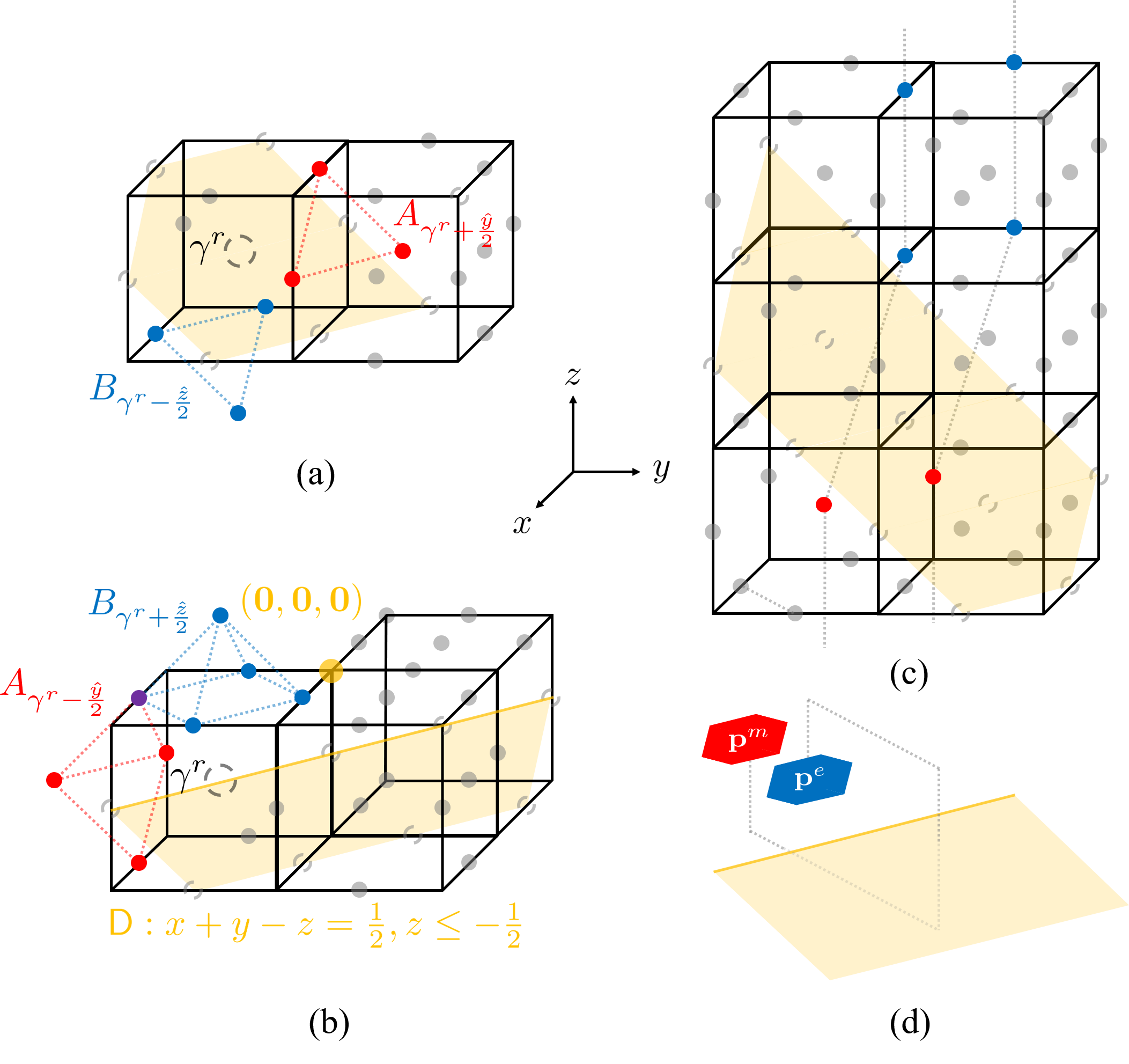}
	\caption{Non-Abelian line defect in 3D orthoplex model. In (a) and (b), we demonstrate modified stabilizers around the removed qubits far from and on the dislocation line, respectively, where the modified stabilizers are obtained by taking the products of the truncated $A$ and $B$ stabilizers. Note that in (a), the modified stabilizer preserve the original orthoplex geometry but is ``stretched''. Qubits involved in $A$, $B$ and both of them are respectively colored red, blue and purple. In (c) and (d), we demonstrate how a planon changes its type (i.e., $\p^e \leftrightarrow \p^m$) when it circles the dislocation line, as it must have crossed the half plane of removed qubits for an odd number of times.}
	\label{fig:defect}
\end{figure}

Finally, we demonstrate that a line defect can manifest non-Abelian statistics by exchanging planon types ($\p^e \leftrightarrow \p^m$). For simplicity, we consider open boundary conditions and focus on the bulk effect of the defect. We introduce a dislocation line along $x+y=0, z=-1/2$ by removing all qubits in the half-plane $\D$ defined by $x+y-z=\frac{1}{2}, z\leq -\frac{1}{2}$ (see Fig.~\ref{fig:defect} (b)). The Hamiltonian must be modified near the cut $\D$. The principle is to pair up truncated stabilizers whose support is partially removed by $\D$ to form new, commuting terms. Specifically, removing a qubit at position $\gamma^r$ truncates the six stabilizers in its neighborhood. These truncated operators are then paired to form new stabilizers by fusing truncated $A$-type and $B$-type stabilizers into mixed-type stabilizers. This pairing rule depends on the location relative to the defect line.

To compensate for the removal of a qubit $\gamma^r$ away from the dislocation line, the new neighboring stabilizers are formed according to its $z$-coordinate. For instance, if the $z$-coordinate of $\gamma^r$ is a half-integer, the new stabilizers are (see Fig.~\ref{fig:defect} (a)):
\begin{align*}
 &A_{\gamma^r+\frac{1}{2} \hat{y}} B_{\gamma^r-\frac{1}{2} \hat{z}},\\ &A_{\gamma^r-\frac{1}{2} \hat{y}} B_{\gamma^r+\frac{1}{2} \hat{z}},
\end{align*}
where new stabilizers involving $A_{\gamma^r \pm \frac{1}{2} \hat{x}}$ are omitted as they can be obtained by considering other removed qubits with the same $z$-coordinate.
 
Note that as the $z$-coordinate of the dislocation line is $-1/2$, a similar pairing scheme applies to removed qubits on the dislocation line, though the paired operators may have different forms (see Fig.~\ref{fig:defect} (b)). Nevertheless, to make resulting stabilizers mutually commute, we need to partition the dislocation line into two sublattices and apply the pairing scheme for only qubits in one sublattice, and delete $A_{\gamma^r-\frac{1}{2} \hat{y}}$ and $B_{\gamma^r+\frac{1}{2} \hat{z}}$ for $\gamma^r$ from the other sublattice. The whole construction removes more stabilizers than qubits and leads to zero modes along the dislocation line~\cite{Bombin2010}, while such zero modes are irrelevant for our discussion here. The crucial consequence of this construction is the emergence of mixed-type stabilizers near $\D$, as shown in Fig.~\ref{fig:defect} (a). When a planon crosses the defect plane, its string operator must commute with these new stabilizers, which, due to the mixed nature of the stabilizers, forces the planon to transform its type (e.g., $\p^e$ becomes $\p^m$). This transformation is observable for paths that encircle the dislocation line, confirming its non-Abelian statistics (see Fig.~\ref{fig:defect} (c) and (d)).

\section{4D Orthoplex model}
\label{sec:4d_model}

In this section, we introduce the 4D orthoplex model and use its excitations as a concrete example to demonstrate the concept of \textit{fragmented topological excitations}.

\begin{figure}[t]
	\centering
	\includegraphics[width=\columnwidth]{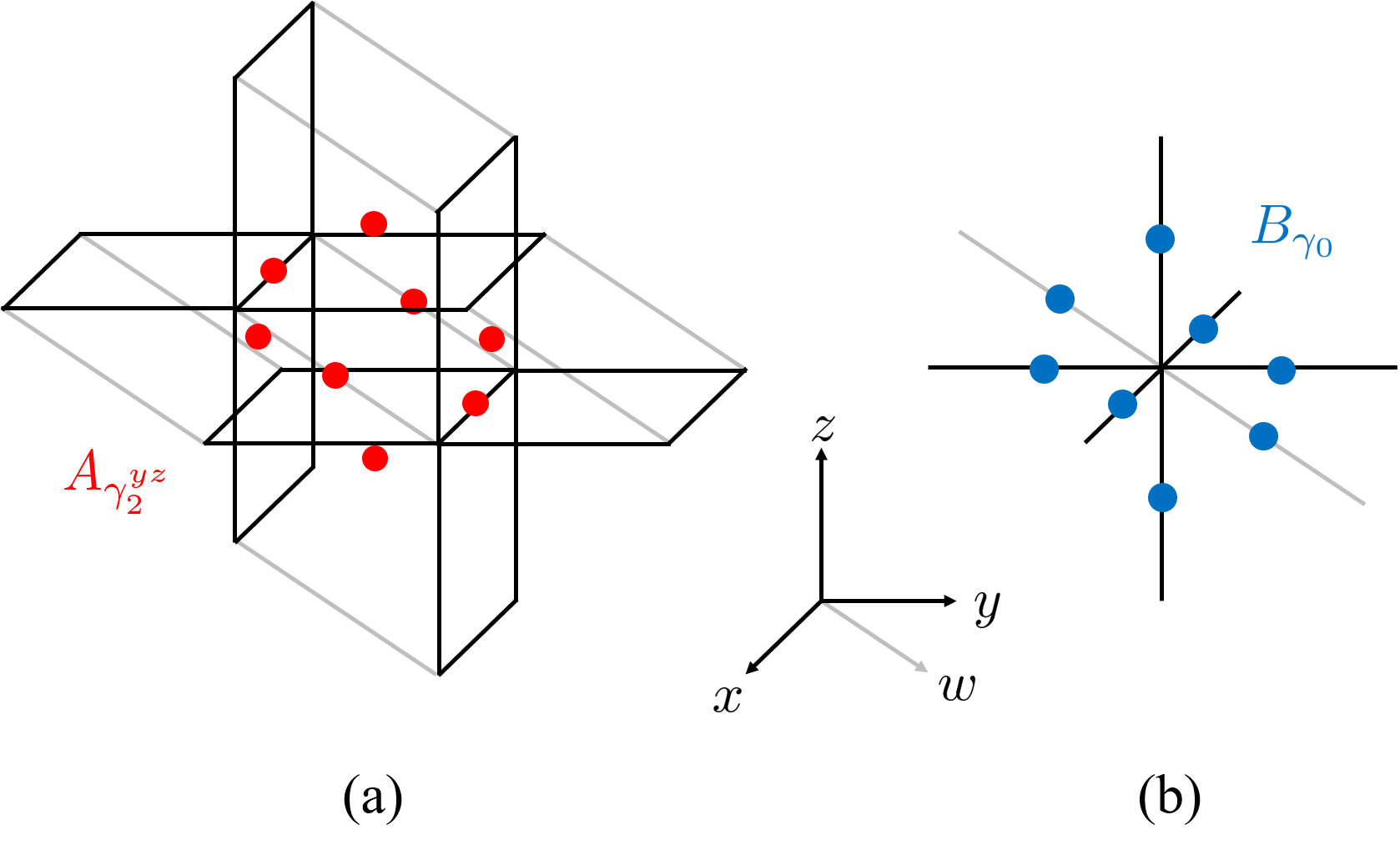}
	\caption{Hamiltonian of 4D orthoplex model. In (a) we show a representative $A_{\gamma^{yz}_2}$ term that involves eight qubits on nearest $\gamma_1$ and $\gamma_3$. In (b), we show a representative $B_{\gamma_0}$ term that involves eight qubits on nearest $\gamma_1$. In both cases, the qubits form vertices of a 4D orthoplex. For clarity, we draw links along direction $w$ by gray lines.}
	\label{fig:4dmodel}
\end{figure}

The 4D orthoplex model is defined on a 4D hypercubic lattice of size $L_x \times L_y \times L_z \times L_w$, with coordinates denoted as $x, y, z,$ and $w$. Qubits and stabilizers are assigned as follows (see Fig.~\ref{fig:4dmodel}):
\begin{itemize}
	\item \textbf{Qubits}: A qubit is assigned to each link ($\gamma_1$) and each cube ($\gamma_3$).
	\item \textbf{$X$-stabilizers}: An $X$-stabilizer is assigned to each plaquette ($\gamma_2$) normal to the $xy$-, $xz$-, or $yz$-planes, and to each hypercube ($\gamma_4$).
	\item \textbf{$Z$-stabilizers}: A $Z$-stabilizer is assigned to each vertex ($\gamma_0$) and each plaquette ($\gamma_2$) normal to the $xw$-, $yw$-, or $zw$-planes.
\end{itemize}

The Hamiltonian is a sum of commuting stabilizer operators:
\begin{align}
	H_\text{4D} = -\sum_{\gamma^s \in {\Gamma_A}} A_{\gamma^s} -\sum_{\gamma^s \in {\Gamma_B}} B_{\gamma^s},
\end{align}
where $\Gamma_A$ denotes the set of $\gamma_2^{xy}$, $\gamma_2^{xz}$,  $\gamma_2^{yz}$ and $\gamma_4$, $\Gamma_B$ denotes the set of $\gamma_0$, $\gamma_2^{xw}$, $\gamma_2^{yw}$ and $\gamma_2^{zw}$, $\gamma_2^{\mu \nu}$ denotes a plaquette normal to the plane spanned by the $\mu$- and $\nu$-axes. The stabilizer operators are defined as $A_{\gamma^s} = \prod_{\gamma^q \in \Delta \gamma^s} X_{\gamma^q}$ and $B_{\gamma^s} = \prod_{\gamma^q \in \Delta \gamma^s} Z_{\gamma^q}$, where the product runs over all qubit-bearing cells $\gamma^q$ in the boundary of the stabilizer cell $\gamma^s$. This relationship implies that each stabilizer involves eight qubits located at the vertices of a 4D orthoplex centered on the stabilizer, analogous to the 3D case discussed in Sec.~\ref{sec:3d_model}. The stabilizers are again layered: 4D orthoplexes centered on 3D hyperplanes with a half-integer $w$ coordinate correspond to $A$ stabilizers, while those on hyperplanes with an integer $w$ coordinate correspond to $B$ stabilizers.

As the $A$ and $B$ terms are physically analogous, we focus on excitations created by Pauli-$X$ operators violating $B$ terms. A single violated $B_{\gamma^s}$ term constitutes a lineon excitation, denoted by $\l^e$, as it can be moved freely along the $w$-direction by applying operators like $X_{\gamma^s\pm \frac{1}{2} \hat{w}}$. Applying $X_{\gamma^s\pm \frac{1}{2} \hat{k}},\ k=x,y,z$ violates $B_{\gamma^s}$ terms inside a 3D hyperplane, which leads to more complicated behavior of excitations as will be demonstrated in this section.

\subsection{Spontaneous string orientation}

\begin{figure}[t]
	\centering
	\includegraphics[width=\columnwidth]{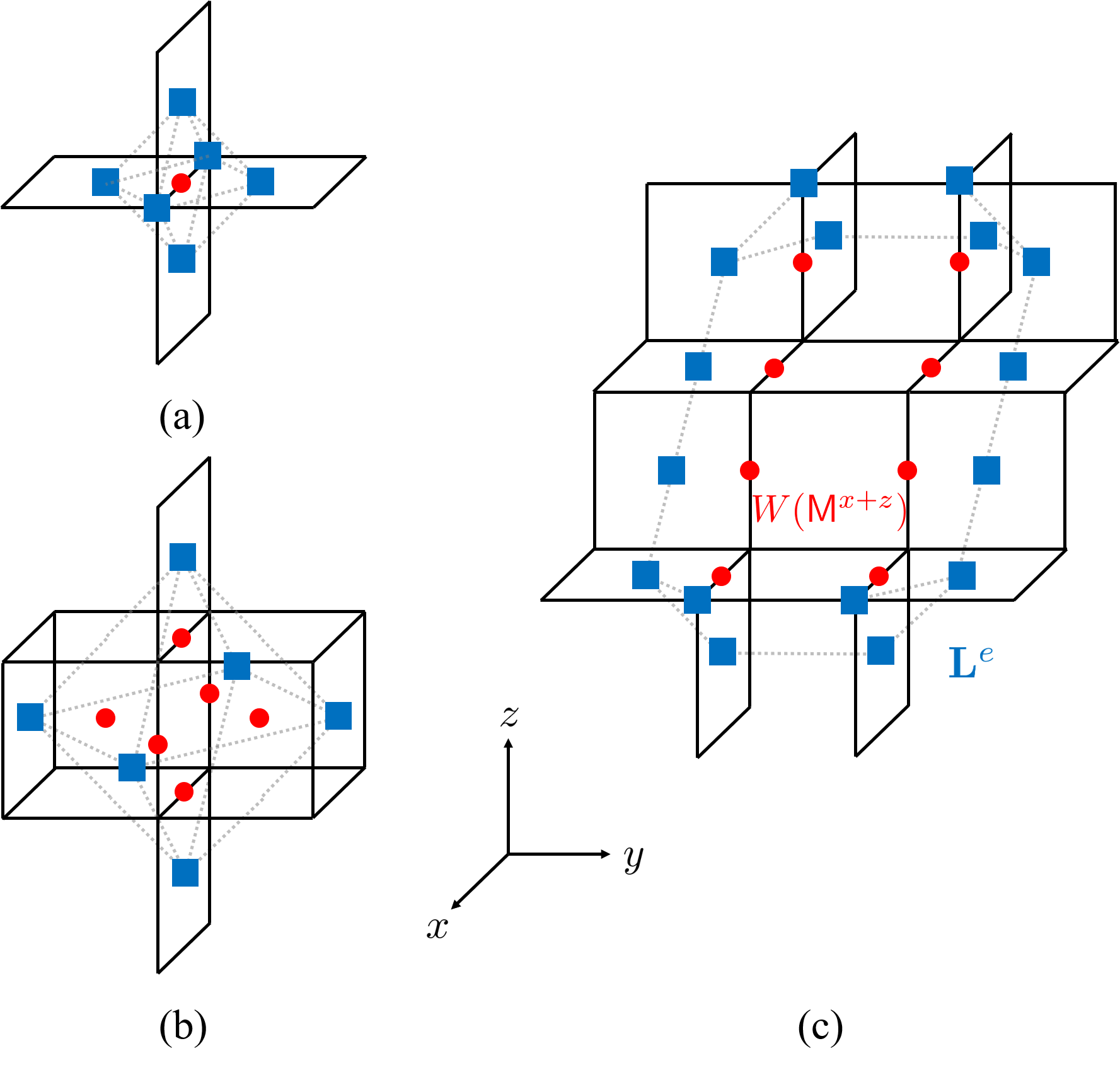}
	\caption{Excitations in a 3D hyperplane of 4D orthoplex model. Qubits acted by Pauli $X$ operators and violated $B$ terms are respectively colored red and blue. In (a), we show the six $B$ terms violated by a single Pauli $X$ operator. In (b), we demonstrate that applying the operator shown in (a) at six vertices of a octahedron also violates six $B$ terms, located on the vertices of a larger octahedron. In (c), we show an $\L^e$ loop excitation created by membrane operator $W(\M^{x+z})$. Note that (a) can also be regarded as the case when $W(\M^{x+z})$ reduces to a single qubit. For this ``minimal membrane'', its orientation is undetermined, as it can extends to be a $W(\M^{x\pm z})$ or $W(\M^{x\pm y})$ membrane operator, which create loop excitations restricted in different subspaces (see Sec.~\ref{subsec:loop_res}).}
	\label{fig:4dloop}
\end{figure} 

In the remainder of this section, we focus on a 3D hyperplane specified by $w=0$. In this hyperplane, a single Pauli-$X$ operator flips the six nearest neighbor $B$ terms (note that in this hyperplane all spins are located on $\gamma^s\pm \frac{1}{2} \hat{k},\ k=x,y,z$, where $\gamma^s$ is the location of a $B$ term). Without loss of generality, consider applying an $X$ operator to a link $\gamma_1^x$ oriented along the $x$-direction. This flips $B$ terms at $\gamma_1^x \pm \frac{1}{2}\hat{x}$ (two vertices) and $\gamma_1^x \pm \frac{1}{2}\hat{y}, \gamma_1^x \pm \frac{1}{2}\hat{z}$ (four plaquettes), as shown in Fig.~\ref{fig:4dloop} (a). Superficially, this cluster of excitations appears to form a closed membrane surrounding $\gamma_1^x$. However, as shown in Fig.~\ref{fig:4dloop} (b), applying Pauli $X$ at six vertices of a octahedron only violates six $B$ terms forming vertices of a larger octahedron. Again, we can apply six such octahedron operators to form a larger octahedron that violates only six $B$ terms. By iterating this action, we can see that such a violated $B$ term, as a lineon $\l^e$, is analogous to a ``fracton'' in the hyperplane, where its mobility is totally restricted~\cite{Haah2011,Vijay2015}.

Naturally, $\l^e$ can also gain mobility inside the 3D hyperplane by forming composites, similar to the planons in 3D orthoplex model. Interestingly, here $\l^e$ can form loop excitations with partially restricted mobility and deformability. To see this, let us consider a specific example. For a membrane $\M^{x\pm z}$ embedded within a plane defined by $x\pm z =C$ (where $C$ is a constant), the membrane operator $W(\M^{x\pm z})=\prod_{\gamma^q \in \M^{x\pm z}} X_{\gamma^q}$ flips $B$ terms only along the boundary of $\M^{x\pm z}$. Thus, $W(\M^{x\pm z})$ creates a loop excitation, denoted by $\L^e$, along the boundary of $\M^{x\pm z}$ (see Fig.~\ref{fig:4dloop} (c)). Similarly, we can define operators $W(\M^{y\pm z})$ and $W(\M^{x\pm y})$, both of which create loop excitations. Notably, when we reduce the support of $\M^{x\pm y}$ or $\M^{x\pm z}$ to a single qubit, the resulting excitations are indistinguishable (see Fig.~\ref{fig:4dloop} (a)). This observation indicates that the orientation of the loop excitation emerges spontaneously upon expansion. This behavior stands in sharp contrast to standard topological codes, such as the 3D and 4D toric orders~\cite{Dennis2002, Hamma2005}. We refer to this phenomenon as \textit{spontaneous string orientation}.

\subsection{Mobility and deformability restrictions}
\label{subsec:loop_res}

\begin{figure}[t]
	\centering
	\includegraphics[width=\columnwidth]{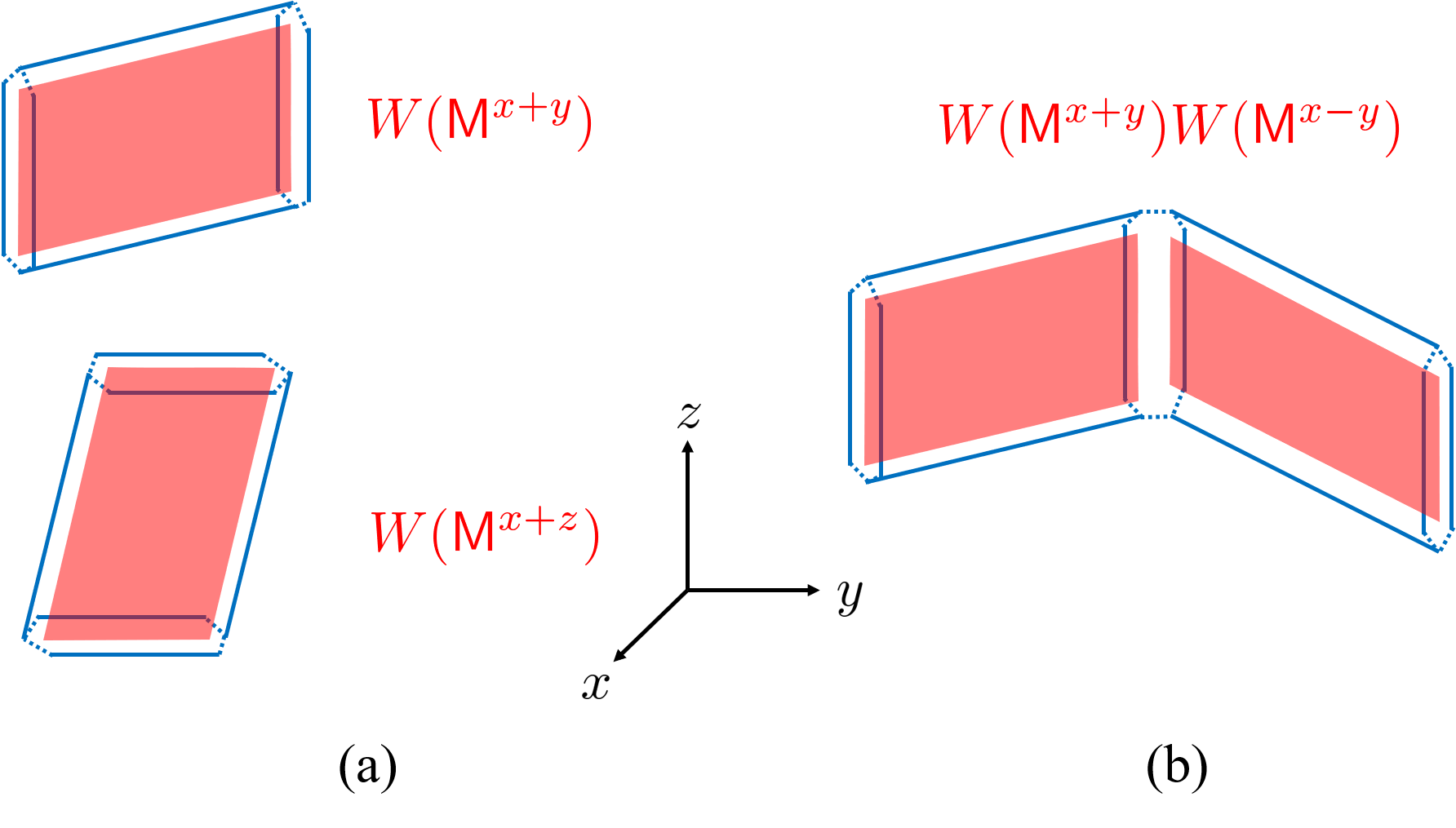}
	\caption{Mobility and deformability restrictions of $\L^e$ loop excitations in 4D orthoplex model. In (a), we show two $\L^e$ excitations created by $W(\M^{x+y})$ and $W(\M^{x+z})$, where energy distribution and supports of creation operators are respectively colored blue and red. Obviously, such $\L^e$ excitations cannot overlap on any continuous line segments. In (b), we demonstrate that a $W(\M^{x-y})$ operator can deform the $\L^e$ excitation created by $W(\M^{x+y})$, but it unavoidably introduces additional energy cost along the intersection of $\M^{x+y}$ and $\M^{x-y}$. From another perspective, we can also regard the excitation in (b) as a whole, that falls into the category of complex excitations characterized by non-manifold-like shapes~\cite{Li2020}.}
	\label{fig:loop_res}
\end{figure}

Next, we demonstrate the restricted mobility and deformability of $\L^e$. Consider a loop $\L^e_1$ created by $W(\M^{x+ y})$ as shown in Fig.~\ref{fig:loop_res} (a). For simplicity, assume $\M^{x+ y}$ lies in the plane $x+y=\frac{1}{2}$ and comprises qubits $\{\gamma^q=(x_1,y_1,z_1) \mid x_1 + y_1 =\frac{1}{2}, 0\leq y_1 \leq l_y, 0\leq z_1 \leq l_z\}$, where $l_y$ and $l_z$ are positive integers. To understand the deformability restriction, we examine the energy density of $\L^e_1$. The loop $\L^e_1$ consists of four segments. Two of them are aligned along diagonal lines where $z_1$ is a constant:
\begin{equation*}
	\begin{split}
		&\L^e_{1,x+y-}:\ x_1 + y_1 = \frac{1}{2}, 0\leq y_1 \leq l_y, z_1 = - \frac{1}{2},\\
		&\L^e_{1,x+y+}:\ x_1 + y_1 = \frac{1}{2}, 0\leq y_1 \leq l_y, z_1 = l_z + \frac{1}{2},
	\end{split}
\end{equation*}
and the other two segments are aligned along the $z$-axis where $x_1$ and $y_1$ are constants:
\begin{equation*}
	\begin{split}
		&\L^e_{1,z-}:\left\{
		\begin{aligned}
			x_1 &= \frac{1}{2}, & y_1 &= -\frac{1}{2}, && 0 \leq z_1 \leq l_z, \\
			x_1 &= 1,           & y_1 &= 0,            && 0 \leq z_1 \leq l_z,
		\end{aligned}
		\right. \\
		&\L^e_{1,z+}:\left\{
		\begin{aligned}
			x_1 &= -l_y,               & y_1 &= l_y,               && 0 \leq z_1 \leq l_z, \\
			x_1 &= -l_y + \frac{1}{2}, & y_1 &= l_y + \frac{1}{2}, && 0 \leq z_1 \leq l_z.
		\end{aligned}
		\right.
	\end{split}
\end{equation*}
We denote these segments as $\L^e_{1,x+y-}$, $\L^e_{1,x+y+}$, $\L^e_{1,z-}$, and $\L^e_{1,z+}$, respectively. Similar to the loop excitation created by $W(\M^{x+z})$ shown in Fig.~\ref{fig:4dloop}, the diagonal segments $\L^e_{1,x+y\pm}$ are composed of flipped $B_{\gamma_0}$ and $B_{\gamma_2^{zw}}$ terms forming a single diagonal line. In contrast, the vertical segments $\L^e_{1,z\pm}$ are composed of flipped $B_{\gamma_2^{xw}}$ and $B_{\gamma_2^{yw}}$ terms that form ``double lines'' of excitations. Consequently, the energy density of the vertical segments $\L^e_{1,z\pm}$ is $\sqrt{2}$ times that of the diagonal segments $\L^e_{1,x+y\pm}$.

This geometry of the segments of loop excitations leads to rigidity. As illustrated in Fig.~\ref{fig:loop_res} (a), $W(\M^{x\pm z})$ creates $\L^e_{1,x\pm z\pm}$ and $\L^e_{1,y\pm}$. Similarly, $W(\M^{y\pm z})$ creates excitation segments $\L^e_{1,y \pm z\pm}$ and $\L^e_{1,x\pm}$. Therefore, neither operator can cleanly annihilate the segments of $\L^e_{1}$ created by $W(\M^{x+ y})$. The only remaining possibility for deformation is to use $W(\M^{x- y})$. However, the vertical segment $\L^e_{1,z\pm}$ created by $W(\M^{x+ y})$ involves $B_{\gamma_2^{xw}}$ and $B_{\gamma_2^{yw}}$ terms separated by a displacement vector $(-\frac{1}{2}, -\frac{1}{2}, 0)$, whereas the corresponding segment created by $W(\M^{x- y})$ involves terms separated by $(\frac{1}{2}, -\frac{1}{2}, 0)$. As a result, as shown in Fig.~\ref{fig:loop_res} (b), even if we apply $W(\M^{x+ y})$ and $W(\M^{x- y})$ to create excitation segments along the same line segment in the $z$-direction—attempting to create a ``curved'' loop along the boundary of the combined membrane $\M^{x+ y}\cup \M^{x- y}$—a residual excitation segment inevitably persists at the crease where $\M^{x+ y}$ and $\M^{x- y}$ meet. This additional energy cost prevents the loop excitation from curving. Alternatively, the object created by $W(\M^{x+ y}) W(\M^{x- y})$ can be viewed as a composite, exotic string-like excitation with a non-manifold-like shape. Such an excitation is dubbed a ``chairon'', serving as a typical example of complex excitations discussed in Ref.~\cite{Li2020}. Barring the formation of such complex excitations, a loop excitation $\L^e$ is energetically constrained to move and deform only within the plane in which its generating membrane $\M$ is embedded. 

\begin{figure}[t]
	\centering
	\includegraphics[width=\columnwidth]{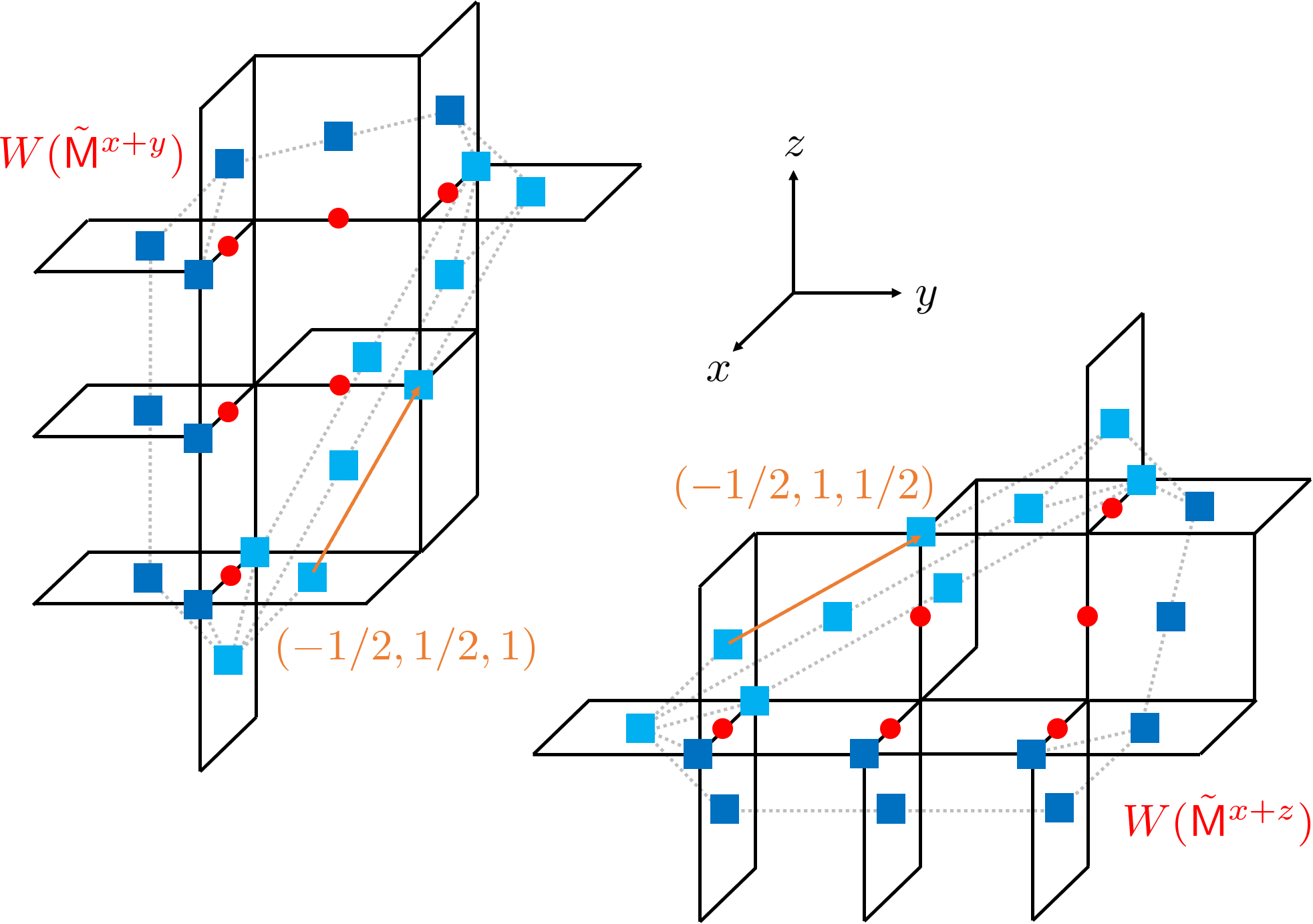}
	\caption{Space diagonal segments of $\L^e$ excitations. Without loss of generality, we illustrate a $W(\tilde{\M}^{x+y})$ and a $W(\tilde{\M}^{x+y})$ operators and the two $\L^e$ loop excitations generated by them. Following the convention in Fig.~\ref{fig:4dloop}, we use red dots and blue squares to respectively denote Pauli $X$ action and flipped $B$ terms. For clarity, flipped $B$ terms composing space diagonal segments are highlighted with light blue. As we can see, the space diagonal segments created by $W(\tilde{\M}^{x+y})$ and $W(\tilde{\M}^{x+z})$ are of different directions, which are respectively along displacement vectors $(-1/2,1/2,1)$ and $(-1/2,1,1/2)$ (denoted by orange arrows). Consequently, such segments cannot have overlap of finite length.}
	\label{fig:loop_diag}
\end{figure}

Moreover, we consider the space diagonal segments of $\L^e$, which can be created by $W(\tilde{\M})$ supporting on triangular membrane $\tilde{\M}$. As demonstrated in Fig.~\ref{fig:loop_diag}, the boundaries along space diagonal directions of triangular $\M$ of different orientations cannot overlap, thus this kind of segments of $\L^e$ cannot violate the mobility and deformability restrictions, either.

Finally, we remark that $\L^e$ in parallel planes are not independent: two $\L^e$ in two adjacent parallel planes can lower their total energy by aligning with each other such that some of the flipped $B$ terms would be canceled. This effect allows a series of delicately assigned $\L^e$ to be reduced to a discrete set of $\l^e$ excitations only mobile along $w$ direction (see Fig.~\ref{fig:4dloop}). This result is consistent with the fact that an $\L^e$ is a composite of $\l^e$.

\subsection{Fragmentation of loop excitations}

Although within a 3D hyperplane of fixed $w$, a loop excitation $\L^e$ is constrained to move and deform within a single 2D plane, it is crucial to recognize that $\L^e$ is microscopically composed of a series of flipped $B$ terms. As established previously, a single flipped $B$ term constitutes a lineon, which can move freely along the $w$-direction. This implies that, fundamentally, the loop excitation $\L^e$ is a collection of lineons.

\begin{figure}[t]
	\centering
	\includegraphics[width=\columnwidth]{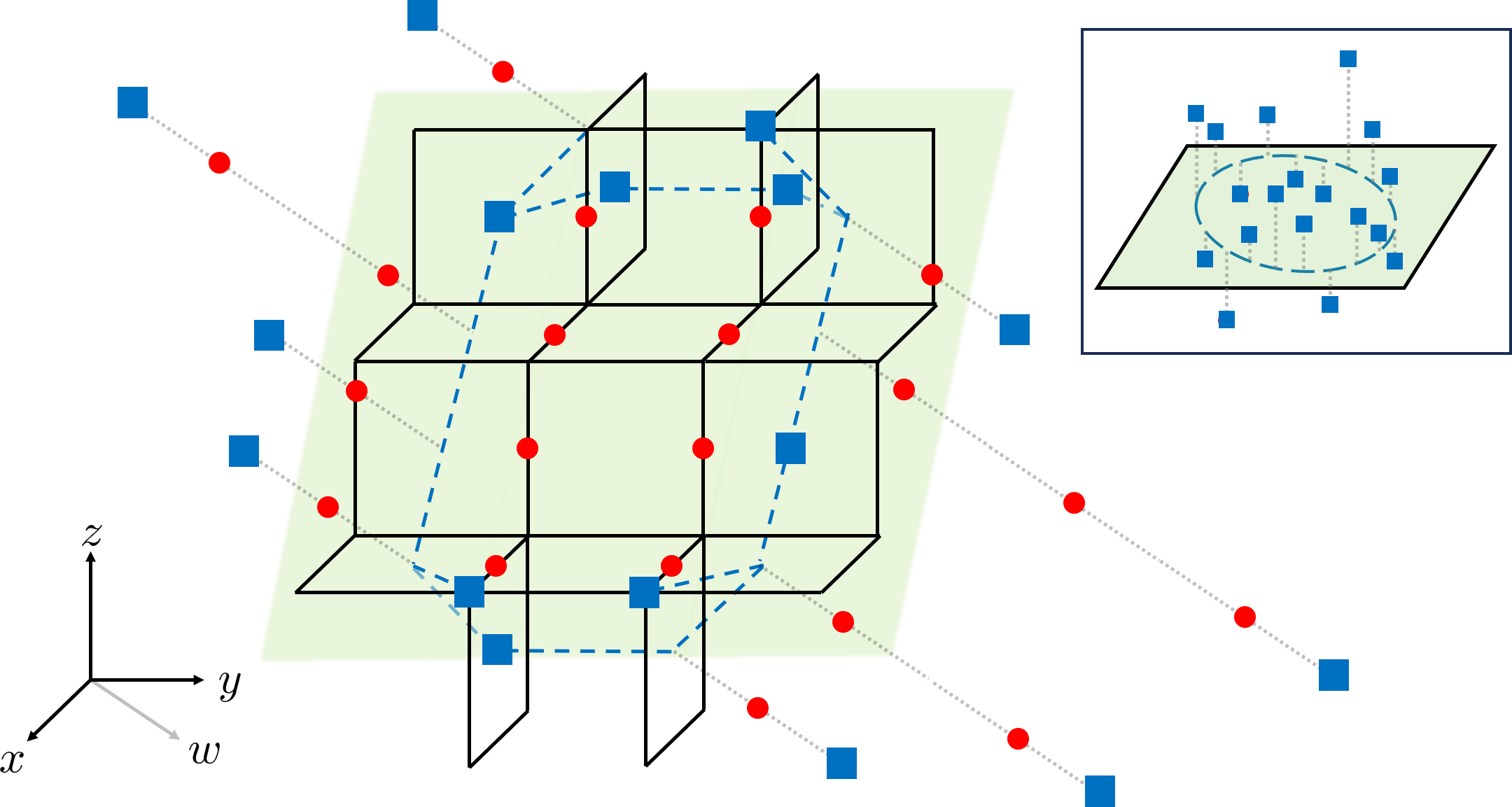}
	\caption{A fragmented loop excitation in 4D orthoplex model. This excitation is obtained based on the $\L^e$ excitation in Fig.~\ref{fig:4dloop} (a). Such an excitation is created by the product of a series of Pauli $X$ operators represented by red dots, and the excitation is composed of flipped $B$ terms represented by blue squares. As we can see, this excitation seems to be composed of scattered quasiparticles, while its projection onto a plane (colored green) forms a connected loop represented by the dashed blue lines. For clarity, the inset shows a schematic picture of this excitation.}
	\label{fig:dis_lat}
\end{figure}

Since each constituent lineon can be independently displaced along the $w$-axis, the valid configuration of an $\L^e$ loop can manifest in 4D real space as a cloud of scattered point-like excitations distributed across a 3D region (spanned by the 2D plane of the loop and the $w$-axis).  Nevertheless, as illustrated in Fig.~\ref{fig:dis_lat}, because such a fragmented configuration is generated by a membrane operator $W(\M)$ followed by $w$-translations of constituent lineons, the projection of these scattered points back onto the plane of $\M$ must retain the topology of a closed loop. This projection reveals the underlying, preserved topological nature of the excitation, despite its spatially disconnected appearance. This peculiar structure renders $\L^e$ an intermediate type between point-like and spatially extended excitations, which we term \textit{fragmented topological excitations}.

Fragmented topological excitations reveal rich physical implications. First, topological orders are generally characterized by their excitation data, composed of both point-like and spatially extended types~\cite{Hamma2005,Walker2012,Lan2018,Lan2019,Chan2018,Wen2018a,Wang2014,Jian2014,Jiang2014,Ning2016,Ning2018a,Ye2018,Wang2015a,Chen2016,Wang2016b,Putrov2017}. Fragmented excitations serve as an intermediate between these types, thereby expanding our general understanding of topological phases. Moreover, although in this paper we explicitly construct fragmented excitations in 4D, this phenomenon may not be exclusive to high-dimensional systems and could potentially exist in 3D systems. Generally, the concept suggests that there may be hidden topological structures governing collections of point-like excitations, offering rich possibilities for revisiting known quasiparticle excitations. Developing theoretical tools to identify such hidden topological structures in systems with only seemingly point-like excitations appears to be a promising direction.

Finally, fragmented spatially extended excitations may play a role in quantum error correction (QEC). In practice, imperfect measurements in QEC codes require repeated measurement cycles to ensure fault tolerance, incurring a time cost. Single-shot codes have been proposed to eliminate this additional cost by leveraging spatially extended syndromes, which provide sufficient redundancy to identify measurement errors within a single round~\cite{Dennis2002, Bombin2009, Bombin2015, Quintavalle2021, Higgott2023}. Fragmented excitations may offer an exotic alternative for this problem: even if the syndromes appear as scattered points, their intrinsic topological structures could provide the necessary information to locate measurement errors and preserve the logical information.

\section{Orthoplex model in arbitrary dimensions}
\label{sec:hd_models}

\subsection{Review of hypergraph product (HGP) codes}
\label{subsec:rev_hgp}

Before introducing the general construction of orthoplex models, we briefly review the original hypergraph product (HGP) codes using the language of chain complexes~\cite{Tillich2009}. This algebraic formalism, which inspires the generalizations introduced in Ref.~\cite{Wu2025} and the exploration in this paper, provides a systematic framework for understanding product codes. We use the term ``HGP codes'' to refer to both the original construction and its arbitrary dimensional generalizations proposed in Ref.~\cite{Zeng2019a}. Note that here, ``dimension'' refers to the number of input classical codes and is not necessarily related to spatial dimensions.

The construction of a $p$-dimensional HGP code begins with $p$ classical linear codes. Each input code is represented by a chain complex of length-$2$ $\mathcal{C}^{i}$ over $\mathbb{Z}_2$, denoted as $C_1^{i} \xrightarrow{\delta^{i}} C_0^{i}$ for $i=1, \dots, p$. Here, $C_0^{i}$ and $C_1^{i}$ are vector spaces over $\mathbb{Z}_2$, with basis elements corresponding to the bits and $Z$-stabilizers of the classical code, respectively\footnote{In general, classical linear codes are specified by classical checks. Here we regard checks as $Z$-stabilizers to consider classical and quantum codes in a unified manner. For clarity, we use ``bit'' for classical codes and ``qubit'' for quantum codes, though classical bits can also be represented by Pauli operators.}. The boundary map $\delta^{i}$ maps a basis element of $C_1^i$ (a $Z$-stabilizer) to the sum of basis elements in $C_0^i$ (a linear combination of bits) on which it acts, thereby defining the action of the stabilizer. We slightly abuse notation by using $\mathcal{C}^{i}$ for both the code and its corresponding chain complex.

The tensor product of these individual complexes yields a product chain complex of length $p+1$, given by $K = \mathcal{C}^{1} \otimes \mathcal{C}^{2} \otimes \dots \otimes \mathcal{C}^{p}$. The chain group of degree $q$ in this product complex is the direct sum
\begin{align}
	K_q = \bigoplus_{i_1 + \dots + i_p = q} \left( C_{i_1}^{1} \otimes \dots \otimes C_{i_p}^{p} \right),
\end{align}
where each $i_j \in \{0, 1\}$. The boundary operator $\partial_q: K_q \to K_{q-1}$ is defined as
\begin{align}
	\partial_q = \sum_{j=1}^{p} I \otimes \dots \otimes \delta^{j} \otimes \dots \otimes I,
\end{align}
where the identity operator $I$ acts on the appropriate component chain groups. Since we only consider codes over $\mathbb{Z}_2$, all chain groups are vector spaces with $\Z_2$ coefficients. Consequently, the sign factors typically required for the boundary operator of a tensor product complex are trivial and have been omitted.

A Quantum CSS code can be constructed from any segment of this complex $K$ of length $3$, namely $K_{q+1} \xrightarrow{\partial_{q+1}} K_{q} \xrightarrow{\partial_q} K_{q-1}$, whose code space is isomorphic to the $q$-th homology group $H_q(K) = \ker(\partial_q) / \mathrm{im}(\partial_{q+1})$. The basis elements of $K_q$, $K_{q+1}$, and $K_{q-1}$ correspond to the physical qubits, $Z$-stabilizers, and $X$-stabilizers of the quantum code, respectively. The boundary operators define the stabilizer actions: $\partial_{q+1}$ maps a $Z$-stabilizer to the sum of qubits it acts upon, and $\partial_q$ maps a qubit to the sum of $X$-stabilizers that include it in their supports. The stabilizer code condition, which requires that the $X$ and $Z$ stabilizers commute, is automatically satisfied due to the fundamental property of a chain complex, $\partial_{q} \circ \partial_{q+1} = 0$: this property ensures that the support of any $Z$-stabilizer has an even intersection with the support of any $X$-stabilizer, guaranteeing their commutation.

This abstract construction admits a more intuitive interpretation. A basis element of $K_q$ can be represented by a $p$-tuple of elements, where exactly $q$ elements are stabilizers and the remaining $p-q$ are bits, each drawn from a distinct input code $\mathcal{C}^i$. Consequently, qubits are represented by $p$-tuples formed by $q$ stabilizers and $p-q$ bits, while $X$- and $Z$-stabilizers are represented by tuples with $q-1$ and $q+1$ stabilizers, respectively. The operator $\partial_{q+1}$ maps a basis element of $K_{q+1}$ (a $Z$-stabilizer) to a sum of basis elements in $K_q$ (qubits) by replacing one of its stabilizer components with the bits on its boundary. Similarly, $\partial_q$ maps a basis element of $K_q$ (a qubit) to a sum of basis elements in $K_{q-1}$ ($X$-stabilizers).

To provide a concrete illustration, we demonstrate the construction of the 3D toric code as an HGP code. The inputs are three identical classical 1D repetition codes, each defined on a periodic chain of length $L$. Such a 1D repetition code, described by the complex $\C: C_1 \xrightarrow{\delta} C_0$, can be viewed as a 1D stabilizer code with only $Z$-stabilizers. The bits and stabilizers are placed on the vertices and links of the 1D chain, respectively, and each stabilizer is a product of Pauli $Z$ operators on two nearest-neighbor bits. This code has a direct correspondence with the 1D Ising model if we regard each stabilizer as a Hamiltonian term. In the chain complex representation, the 0-chain group $C_0$ is spanned by vertices (0-cells) at integer coordinates $\{j | j \in \mathbb{Z}_L\}$, representing bits. The 1-chain group $C_1$ is spanned by links (1-cells) at half-integer coordinates $\{j+1/2 | j \in \mathbb{Z}_L\}$, representing $Z$-stabilizers. The boundary map $\delta: C_1 \to C_0$ is defined by $\delta(j+1/2) = j + (j+1)$, where the right-hand side is the sum of the two basis elements of $C_0$ at the endpoints of the link denoted by $j+1/2$.

The resulting 3D code is constructed from the segment $K_2\xrightarrow{\partial_2} K_1 \xrightarrow{\partial_1} K_0$ of the product complex $K = \mathcal{C} \otimes \mathcal{C} \otimes \mathcal{C}$. As the input codes are identical, we omit their superscripts, which can be inferred from their position in the tensor product. The basis elements of the $q$-chain group $K_q$ correspond to $q$-cells in a 3D cubic lattice of size $L \times L \times L$, thus we have:
\begin{itemize}
	\item \textbf{Qubits ($K_1$):} The chain group $K_1 = (C_1 \otimes C_0 \otimes C_0) \oplus (C_0 \otimes C_1 \otimes C_0) \oplus (C_0 \otimes C_0 \otimes C_1)$ is spanned by basis elements corresponding to the links (1-cells) of the 3D lattice. For example, a basis element $(j_1+1/2) \otimes j_2 \otimes j_3$ corresponds to an $x$-oriented link at coordinate $(j_1+1/2, j_2, j_3)$.
	\item \textbf{$Z$-stabilizers ($K_2$):} The group $K_2 = (C_1 \otimes C_1 \otimes C_0) \oplus (C_1 \otimes C_0 \otimes C_1) \oplus (C_0 \otimes C_1 \otimes C_1)$ is spanned by basis elements corresponding to the plaquettes (2-cells). For instance, $(j_1+1/2) \otimes (j_2+1/2) \otimes j_3$ corresponds to a plaquette perpendicular to direction $z$ at coordinate $(j_1+1/2, j_2+1/2, j_3)$. The operator $\partial_2$ maps each plaquette to the four qubits on its boundary.
	\item \textbf{$X$-stabilizers ($K_0$):} The group $K_0 = C_0 \otimes C_0 \otimes C_0$ is spanned by basis elements corresponding to the vertices (0-cells). A basis element $j_1 \otimes j_2 \otimes j_3$ corresponds to the vertex at integer coordinate $(j_1, j_2, j_3)$. The operator $\partial_1$ maps each qubit to the two vertices at its endpoints.
\end{itemize}
This construction precisely yields the 3D toric code, where qubits reside on links, $Z$-stabilizers on plaquettes, and $X$-stabilizers on vertices.

\subsection{Orthoplex models from generalized hypergraph product codes}
\label{sec:models}

In this section, we demonstrate how to construct \textit{orthoplex models}, a family of quantum codes exhibiting fracton orders, by applying a generalized HGP protocol, which is a subclass of the general code construction introduced in Ref.~\cite{Wu2025}. This protocol extends the standard HGP framework reviewed in Sec.~\ref{subsec:rev_hgp}.

The standard HGP protocol assigns qubits and stabilizers to a segment of length-$3$ of the entire tensor product complex $K = \C^{1} \otimes \dots \otimes \C^{p}$. A key observation is that these chain groups possess a natural direct sum structure. The generalized protocol leverages this structure by treating each direct summand individually. This method moves beyond the conventional homological construction of product codes and enables the creation of diverse quantum codes, including the orthoplex models central to this work.

\begin{figure}[t]
	\centering
	\includegraphics[width=\columnwidth]{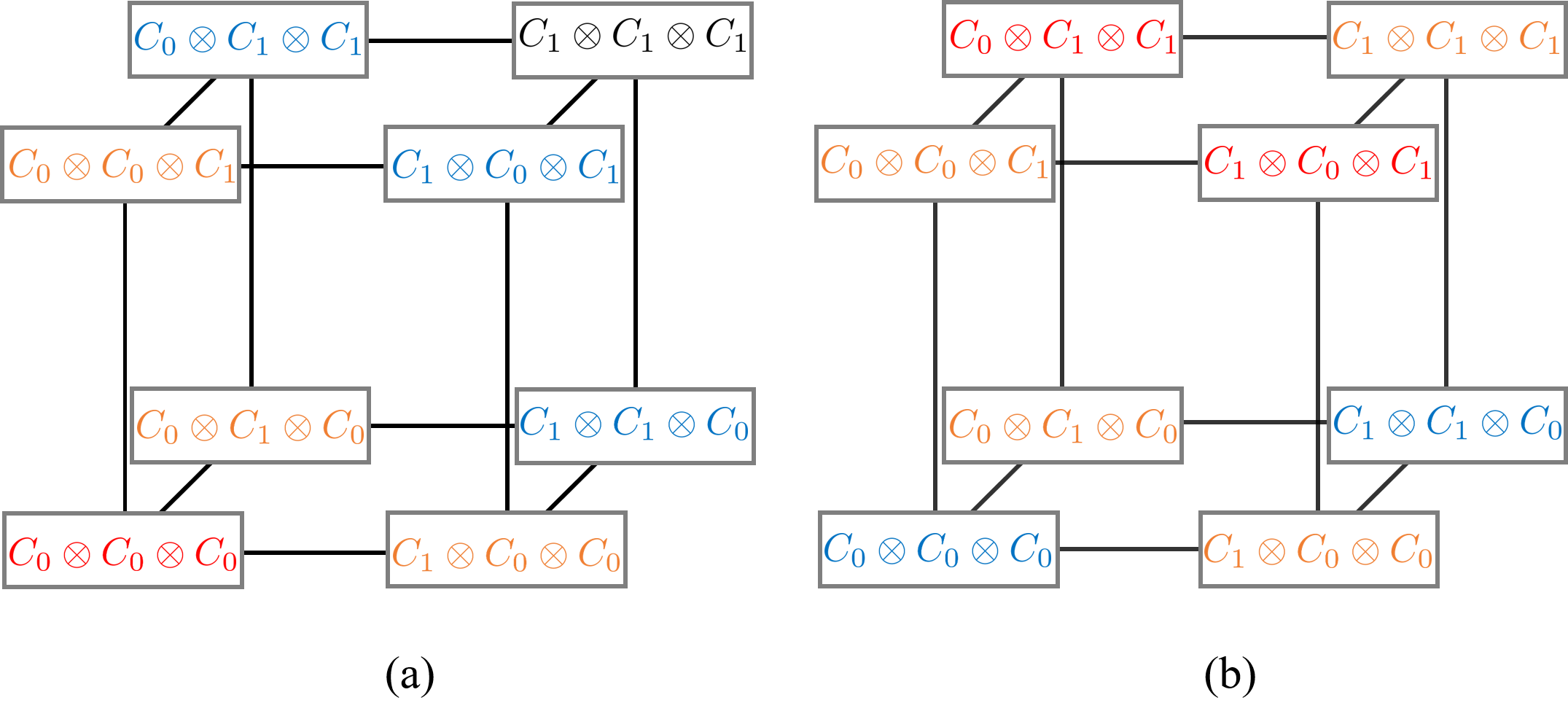}
	\caption{A comparison between the construction of 3D toric code (a) and orthoplex models (b). We visualize the direct summands of chain groups of $K$ as vertices of a cube, and use colors to represent the partitioning of them. More specifically, orange, red, blue and black vertices respectively represent elements of $\mathcal{S}_\text{Q}$, $\mathcal{S}_\text{X}$, $\mathcal{S}_\text{Z}$, and $\mathcal{S}_\text{U}$. Both cases exemplify the general construction protocol of CSS codes in Ref.~\cite{Wu2025}.}
	\label{fig:cube_rep}
\end{figure}

Let us formalize this generalization. The chain group $K_q$ is a direct sum of subspaces of the form $C_{i_1}^{1} \otimes \dots \otimes C_{i_p}^{p}$, where $i_j \in \{0,1\}$ and $\sum_{j=1}^p i_j = q$. In the standard HGP construction, one selects an integer $q$ (typically $1 \le q \le p-1$) and assigns the roles of qubits, $X$-stabilizers, and $Z$-stabilizers to the entire chain groups $K_q$, $K_{q-1}$, and $K_{q+1}$, respectively. The generalization constitutes partitioning the complete set of all direct summands $\{ C_{i_1}^{1} \otimes \dots \otimes C_{i_p}^{p}\}$ into four disjoint sets, denoted as $\mathcal{S}_\text{Q}$, $\mathcal{S}_\text{X}$, $\mathcal{S}_\text{Z}$, and $\mathcal{S}_\text{U}$ (e.g., see Fig.~\ref{fig:cube_rep}). The basis elements of summands in these sets are designated as qubits, $X$-stabilizers, $Z$-stabilizers, or are left unused, respectively. The corresponding chain groups are then defined as $C_\text{Q} = \bigoplus_{\mathcal{V} \in \mathcal{S}_\text{Q}} \mathcal{V}$, $C_\text{X} = \bigoplus_{\mathcal{V} \in \mathcal{S}_\text{X}} \mathcal{V}$, and $C_\text{Z} = \bigoplus_{\mathcal{V} \in \mathcal{S}_\text{Z}} \mathcal{V}$. The boundary operators are defined by combining boundary operators of input codes $\delta^i$, yielding maps $\partial_\text{X}: C_\text{Q} \to C_\text{X}$ and $\partial_\text{Z}: C_\text{Z} \to C_\text{Q}$ that can correctly connect these chain groups. For appropriate partitions of $\{ C_{i_1}^{1} \otimes \dots \otimes C_{i_p}^{p} \}$, the above construction gives well-defined chain complexes satisfying $\partial_X \partial_Z = 0$ such that the CSS condition is naturally satisfied~\cite{Wu2025}. In this work, we focus on a specific partition strategy that leads to orthoplex models, which is defined as follows:

\begin{itemize}
	\item \textbf{Qubits}: The chain group spanned by qubits is $C_\text{Q}=\bigoplus_{i_1, \dots, i_p} (C^1_{i_1} \otimes \cdots \otimes C^p_{i_p})$ where $\sum_j i_j$ is odd.
	\item \textbf{$X$-stabilizers}: The chain group spanned by $X$-stabilizers is $C_\text{X}=\bigoplus_{i_1, \dots, i_p} (C^1_{i_1} \otimes \cdots \otimes C^p_{i_p})$ where $\sum_j i_j$ is even and $i_p=1$.
	\item \textbf{$Z$-stabilizers}: The chain group spanned by $Z$-stabilizers is $C_\text{Z}=\bigoplus_{i_1, \dots, i_p} (C^1_{i_1} \otimes \cdots \otimes C^p_{i_p})$ where $\sum_j i_j$ is even and $i_p=0$.
	\item \textbf{Stabilizers-Qubit Relation}: The boundary operators $\partial_X^T$ and $\partial_Z$ are obtained by restricting the action of full operator $(\bigoplus_{j} I \otimes \dots \otimes \delta^{j} \otimes \dots \otimes I) \oplus (\bigoplus_{k} I \otimes \dots \otimes (\delta^{k})^T \otimes \dots \otimes I)$ to the corresponding domains (i.e., $C_X$ or $C_Z$).
\end{itemize}

The orthoplex models are defined within this general framework, using $p$ identical 1D repetition codes as inputs. As established in Sec.~\ref{subsec:rev_hgp}, the basis elements of the chain groups of the product complex correspond to cells in a $p$-dimensional cubic lattice. A direct summand corresponds to the set of all cells of a given dimension and orientation. The orthoplex models are specified by the following assignment rule:
\begin{itemize}
	\item \textbf{Qubits}: A qubit is assigned to each cell $\gamma_d$ of odd dimension $d$. 
	\item \textbf{$X$-stabilizers}: An $X$-stabilizer is assigned to each cell $\gamma_d$ of even dimension $d$ that extends along the $p$-th dimension.
	\item \textbf{$Z$-stabilizers}: A $Z$-stabilizer is assigned to each cell $\gamma_d$ of even dimension $d$ that does not extend along the $p$-th dimension.
	\item \textbf{Stabilizers-Qubit Relation}: Each stabilizer operator acts on all nearest qubits on the lattice. 
\end{itemize}

The origin of the name ``orthoplex models'' becomes clear from the geometry of the stabilizers. Let us consider an $X$-stabilizer operator. According to the construction, it is associated with an even-dimensional cell, $\gamma_d$. This stabilizer operator acts on the nearest qubits, which reside on odd-dimensional cells $\gamma_{d'} \in \Delta \gamma_d$ located on positions $\gamma_d \pm \frac{1}{2}\hat{x}_\mu$, where $\hat{x}_\mu$ for $\mu=1, \dots, p$ are the basis elements of the lattice. The locations of the $2p$ qubits involved in the stabilizer operator are precisely the vertices of a $p$-dimensional orthoplex (also known as a cross-polytope) centered on the stabilizer operator. For $p=3$, this structure is the familiar octahedron, where six qubits form its vertices. Since a similar geometric argument applies to $Z$-stabilizers, this structure is a fundamental feature of the whole model family, justifying the name ``orthoplex models'' in arbitrary dimensions.

\section{Summary and outlook}
\label{sec:summary}

In this work, we constructed a family of exactly solvable spin models, dubbed orthoplex models, based on a generalized HGP protocol~\cite{Tillich2009,Wu2025}. We investigated the fracton orders within this model family, characterizing features such as non-monotonic GSD, non-Abelian defects, and spatially extended excitations with restricted mobility and deformability. Remarkably, we found that the loop excitations in the 4D orthoplex model exhibit spontaneous orientation and can be ``fragmented'' within a 3D region. Specifically, a fragmented loop excitation manifests as a collection of scattered quasiparticle excitations, yet the projection of these quasiparticles onto a 2D plane preserves the topology of a loop. Such fragmented topological excitations serve as an intermediate class between point-like and spatially extended excitations. Therefore, we expect them not only to play an important role in the characterization of topological orders and the exploration of novel QEC codes, but also to reveal interesting hidden physics in systems superficially hosting only quasiparticle excitations.

Our results naturally suggest several avenues for future research. For example, a systematic classification of fragmented topological excitations would be beneficial for a more complete understanding of topological orders~\cite{Kong2014,Wen2015,Lan2018,Ning2018,Lan2019,Zhang2021,Huang2025}. Furthermore, the potential existence of fragmented topological excitations in 3D may require us to revisit certain familiar quantum many-body systems which seemingly host only point-like excitations. Finally, we remark that the generalized HGP protocol leading to orthoplex models represents only one branch of the general construction of codes in Ref.~\cite{Wu2025}; further exploration of this framework may lead to more novel physics.

\section*{Acknowledgements}
We thank for the beneficial discussion with Hui Zhai, Chengshu Li, Shang Liu, Liang Mao, Yifei Wang, Yu-An Chen and Hao Song. This work is supported by the Shanghai Committee of Science and Technology (Grant No.~25LZ2600800) and  M.-Y. Li is supported by the Shuimu Tsinghua Scholar Program.

\end{document}